\documentclass[runningheads]{llncs}
\usepackage{graphicx}
\usepackage{amsmath}
\usepackage{subfigure}
\usepackage{amsfonts}

\begin{document}
\title{The Cost of Permissionless Liquidity Provision in Automated Market Makers}
\author{Julian Ma \and Davide Crapis}
\authorrunning{J. Ma and D. Crapis}
\institute{Robust Incentives Group, Ethereum Foundation\\
\email{\{julian.ma,davide.crapis\}@ethereum.org}}
\maketitle              % typeset the header of the contribution
\begin{abstract}
    Automated market makers (AMMs) allocate fee revenue \textit{proportional} to the amount of liquidity investors deposit. In this paper, we study the economic consequences of the competition between passive liquidity providers (LPs) caused by this allocation rule. We employ a game-theoretic model in which $N$ strategic agents optimally provide liquidity and two types of liquidity traders trade. In this setting, we find that competition drives LPs to provide excess liquidity. Excess liquidity is costly as more capital is exposed to adverse selection costs. One of our main results is that the price of anarchy, defined over the liquidity provider performance, is $O(N)$, implying that the welfare loss scales linearly with the number of liquidity providers. This inefficient capital allocation is masked when considering the welfare of elastic liquidity traders as the total price of anarchy is $O(1)$. Since this result is driven by elastic liquidity traders benefiting from the liquidity provided because of inelastic liquidity traders, we show that different types of liquidity traders complement each other. Finally, we show that AMM designs that reduce the arbitrage intensity per unit of liquidity do increase utility for liquidity traders but importantly not for LPs nor do they necessarily decrease total arbitrage volume.

\end{abstract}

\newpage
\section{Introduction}

Automated Market Makers (AMMs) are one of the most used applications on blockchains. AMMs allow users to exchange tokens by trading with a smart contract on a blockchain. Liquidity Providers (LPs) deposit tokens into the liquidity pool of the AMM. LPs are passive and do not actively quote prices; instead, prices are determined by a predefined formula based on the current liquidity in the pool, referred to as the bonding curve. A distinct advantage of AMMs over, for example, electronic limit order books, which are the dominant trading mechanism in traditional finance, is that AMMs minimize the computational cost of token exchange. This is important because computation and storage in blockchains are costly. Moreover, AMMs are well-suited for illiquid assets since they do not require a dedicated, active market maker.\\
\indent A downside of AMMs is that passive LPs cannot update their quotes like in traditional finance. New information flows into the market via informed traders trading against the liquidity that the AMM holds in its liquidity pool. These adverse selection costs that passive LPs face in AMMs, known as \emph{Loss-Versus-Rebalancing} (LVR) \cite{milionis2024lvr}, are a big problem for decentralized finance and recent AMM design focuses on mitigating LVR. In this work we highlight an additional factor contributing to the amount of LVR LPs face. Since the fee revenues of an AMM are allocated proportionally to liquidity providers based on their share of the total liquidity, known as the \emph{pro-rata allocation rule}, competition between passive liquidity providers emerges to capture a larger share of the total fee revenue. Therefore, the liquidity providers collectively deposit more liquidity than optimal, which causes a loss in welfare due to more adverse selection costs.\\
\indent In this work, we model the competition between passive liquidity providers. There are $N$ symmetric liquidity providers, a representative elastic liquidity trader and a representative inelastic liquidity trader. The liquidity traders trade for liquidity reasons only and have no price-sensitive information. We use a $N$ player simultaneous subgame with symmetric agents to model liquidity provision. Each LP receives an endowment and decides what fraction to provide as liquidity in an AMM and what fraction to invest in an exogenous investment opportunity. The investments are fixed for a finite time horizon, during which liquidity traders and arbitrageurs trade against the liquidity present in the AMM. Traders pay fees that are realized as income to the liquidity providers. The costs of liquidity provision are adverse selection costs and the opportunity costs of capital.\\
\indent The two types of liquidity traders behave differently. The representative elastic liquidity trader receives a private valuation of the risky asset and trades to maximize its utility. Therefore, we can incorporate the utility of this type of liquidity trader into the price of anarchy of the system. On the other hand, we assume that the representative inelastic liquidity trader trades a fixed amount regardless of trading costs. Since the utility function of this type of trader is outside of the model, we cannot incorporate its utility into the price of anarchy of the system. We purposefully do not endogenize the utility of the inelastic liquidity trader as the reasons for its inelasticity could be very different. The inelastic liquidity trader could be an elastic liquidity trader but whose optimal trading volume is capped by a budget or regulatory constraint. Another reason could be that the inelastic liquidity trader wants to trade on the AMM regardless of price because information acquisition on what the best exchange is is costly, or the trader is locked into a certain system.\\
\indent We show that the amount of liquidity provided because of competition grows linearly with the number of strategic agents and converges to a constant in the limit. This amount of liquidity is also linearly increasing in base demand. One of our main results shows that the price of anarchy in the liquidity provision subgame, defined as the ratio of total profits from the AMM in the case of a single monopolist or a cooperative of agents over the total profits from the AMM in the case of $N$ strategic agents, is $O(N)$. This shows that the welfare loss due to competition between passive liquidity providers scales linearly with the number of liquidity providers. Since providing liquidity is permissionless in decentralized finance, in contrast to traditional finance, there are many LPs, and thus, a significant loss in welfare. The excess liquidity due to competition also creates excess LVR, which has negative externalities on the network \cite{pbs}. On the other hand, the excess liquidity also means that more trade volume settles through the AMM. Our other main result is that the price of anarchy of the liquidity providers and the representative elastic liquidity trader is $O(1)$.\\
\indent These results show that the total amount of liquidity is not an appropriate proxy for welfare for all market participants. Liquidity providers do not necessarily benefit from more liquidity. Furthermore, we show that different types of liquidity traders complement each other. Because there is inelastic demand, there is more liquidity provided to capture the fee revenue. The elastic demand benefits from this additional liquidity.\\
\indent Moreover, we show that a reduction in the arbitrage intensity per unit of liquidity leads to a less than proportional reduction in the total amount of arbitrage, and in some special cases even to an increase, because the equilibrium amount of liquidity deposited grows as the arbitrage intensity per unit of liquidity decreases. These results suggest that AMM designs intended to mitigate LVR do not necessarily decrease the amount of Maximum Extractable Value (MEV) \cite{daian2020flashboys}, of which LVR is an example. Therefore, MEV mitigation in applications may not be sufficient to reduce the centralization of block builders in the protocol layer.\\
\indent The remainder of the paper is structured as follows. In Section 2, the setup of the model is explained. In the following section, the sequential game is layed out in detail with all market participants. In Section 4, we arrive at the main results, including equilibrium liquidity supply and the price of anarchy of the subgame of liquidity provision and of the full game. Finally, we also show that a decrease in arbitrage intensity per unit of liquidity leads to a less-than-proportional reduction in total arbitrage intensity or even an increase.

\subsection{Literature Review}

Our paper contributes to the strand of literature that investigates factors contributing to trading costs of liquidity traders. \cite{capponi2021adoption} show that the curvature of the AMM is a key design consideration as it affects the price impact of swaps performed by arbitrageurs and liquidity traders.\\
\indent \cite{hasbrouck2022need} show that the fees that AMMs charge traders lead to higher levels of liquidity and may, thus, decrease the costs of trading.\\
\indent \cite{hasbrouck2023economic} provide an economic model for liquidity provision in concentrated liquidity pools. Like these papers, our work models the demand for trading of elastic liquidity traders based on certain design considerations of AMMs. However, to the best of our knowledge, this paper is the first to study the effect of the pro-rata allocation rule on trading costs for elastic liquidity traders.\\
Perhaps \cite{capponi2024jit} is most closely related to our work. The authors use a sequential model of liquidity provision with two types of LPs to show that active LPs with the "last look" impose adverse selection costs on passive LPs, yet sometimes these types of LPs complement each other. The results are partially driven by the pro-rata allocation rule. Our paper uses a simultaneous model with two types of liquidity traders. We show that these liquidity traders complement each other. The elastic liquidity trader used in this paper is modeled after the liquidity traders used in \cite{capponi2024jit}.\\
\indent Simultaneously with our paper, \cite{adams2024amamm} develop a new AMM design with a similar model.\\
\indent \cite{johnson2023concave} investigate concave pro-rata games. The subgame of liquidity provision is such a game and we show that the literature extends to this setting when using the model of liquidity traders as proposed in \cite{capponi2024jit}. The full game presented in this paper is not a concave pro-rata game.\\
\indent The model proposed in this paper builds on \cite{milionis2024lvr} and \cite{milionis2024fees}, which find closed-form solutions to the adverse selection costs in AMMs with passive LPs, referred to as \emph{Loss-Versus-Rebalancing} (LVR). Since the focus of our model is on the allocation rule of fee income, we employ the characterization of adverse selection costs proposed in these papers to model liquidity provision endogenously.\\
\indent Finally, our work contributes to the literature on LVR mitigation. The most recent developments in AMM design, such as \cite{canidio2024arbitrageurs}, \cite{mcamm}, and \cite{mcmenamin2022diamonds}, aim to decrease the arbitrage intensity per unit of liquidity. From the liquidity provider perspective, reducing LVR seems favorable as the costs of providing a marginal unit of liquidity decreases. We show, however, that risk-neutral LPs do not gain in equilibrium when the arbitrage intensity per unit of liquidity decreases. Decreasing the arbitrage intensity per unit of liquidity also seems promising to a protocol as MEV, first described by \cite{daian2020flashboys} of which LVR is a specific example, could decrease. We show, however, that a reduction of the arbitrage intensity per unit of liquidity leads to a less than proportional decrease, and in some special cases an increase, in the total amount of arbitrage. Therefore, more research is needed investigating whether MEV mitigation at the application layer can also decrease problems related to MEV extraction at the protocol layer.

\section{Model Setup}

Our starting point is the model of \cite{milionis2024fees}. We assume blocks are generated by a Poisson process with rate $\lambda > 0$ and we consider a time interval of $[0,T]$.

Consider an AMM trading a risky asset, $X$, versus the \textit{numéraire}, $Y$. The fundamental value of the risky asset at time $t$ is given as $p_{t}$, which follows a geometric Brownian motion with volatility $\sigma > 0$, drift $\mu = 0$, and risk-free rate $r > 0$. Let the deposits of the AMM by given by $(x,y)$. We assume the AMM uses the constant function market maker used by Uniswap v2 $\sqrt{xy} = L$, where $L$ is the invariant. We assume the AMM charges a proportional trading fee $f$ that is realized as a separate cashflow to liquidity providers and is denoted in terms of the numéraire.

\section{Sequential Game}

\begin{figure}
    \centering
    \includegraphics[width=\textwidth]{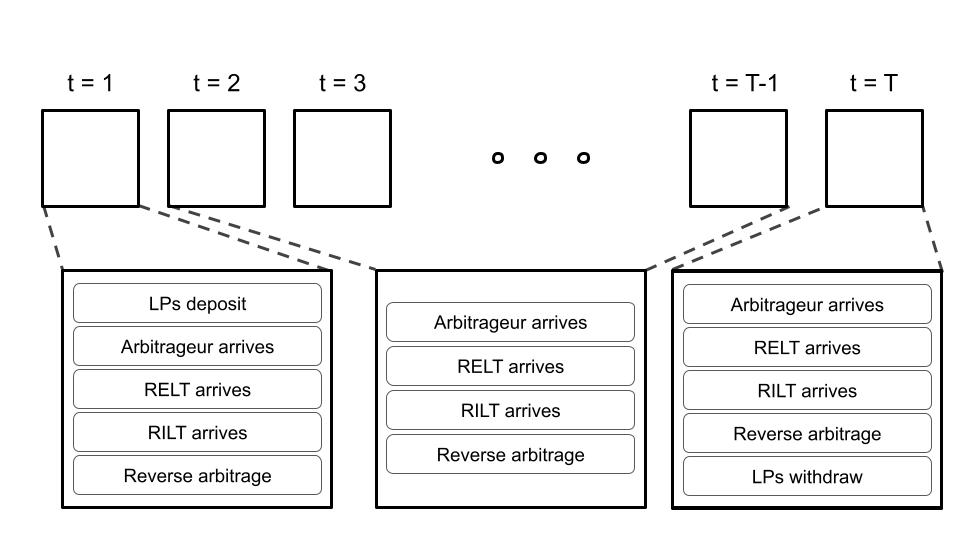}
    \caption{Example of the $T$-period game in which liquidity providers deposit in the first period and withdraw in the last and in which arbitrageurs, representative elastic and inelastic liquidity traders trade in every period.}
    \label{fig:sequential_game}
\end{figure}

We model a $T$-period game in which blocks are generated with a Poisson process. In each block, multiple types of traders trade. which multiple types of traders trade. In the first period, liquidity providers deposit liquidity, and in the final period, they withdraw their liquidity. The liquidity providers and the representative elastic liquidity trader behave as the passive liquidity providers and the liquidity traders of \cite{capponi2024jit}. The full game this paper studies is shown in Figure \ref{fig:sequential_game}.

\subsection{LP Deposits and Withdrawals}
There are $N$ liquidity providers who each receive an endowment of $(e_{X}, p_{0}e_{Y})$. Each LP decides how much assets of their endowment to deposit into the AMM. Let $d_{i} \in [0,e_{X}]$ denote the deposits of risky asset by the $i$th LP into the AMM, then $p_{0}d_{i} \in [0,p_{0}e_{Y}]$ is the deposits of \textit{numéraire}. Let $d = \sum_{i = 1}^N d_{i}$ be the total deposits of risky assets in the pool. 

\subsection{Arbitrageurs}
In each block, an arbitrageur arrives. The arbitrageur behaves as specified in \cite{milionis2024fees}. The arbitrageur maximizes profits and has a perfect signal of the fundamental price of the risky asset. Arbitrageurs may obtain this signal because they monitor a centralized exchange that trades continuously during the time interval. Arbitrageurs decide on $(q_{X}, q_{Y}) \in \mathbb{R}_{+}\backslash \mathbb{R}_{++}$ and they have sufficient amounts of assets of both risky asset and \textit{numéraire}. We assume arbitrageurs do not pay trading fees. This assumption makes the analysis substantially easier yet it does not qualitatively change the results \cite{milionis2024fees}, \cite{milionis2024lvr}. Moreover, some new AMM designs do not require the arbitrageur to pay fees \cite{adams2024amamm}

\subsection{Liquidity Traders}
In our model, there are two types of liquidity traders. These traders trade for liquidity or convenience reasons only and have no information about the future price of the risky asset. A liquidity trader may trade on the AMM because it values trading on-chain or because it must hedge its exposure to the risky asset.

\subsubsection{Representative Elastic Liquidity Trader (ET)}
The representative elastic liquidity trader represents the liquidity traders who trade more when trading costs are lower. Keeping all other factors constant, trading costs become lower when there is more liquidity in the pool because users face less slippage. The representative elastic liquidity trader has no price-relevant information, but it has a private valuation $p_{U}$ for the risky asset.
$$
p_{U} = \begin{cases}
    \gamma^{-1}_{U}p & \text{with probability} \enskip \frac{1}{2} \\
    \gamma_{U}p & \text{with probability} \enskip \frac{1}{2} \\
\end{cases}
$$
with $\gamma_{U} \in (1 + f, \infty)$. The representative elastic liquidity trader trades with the AMM in order to maximize its utility by deciding $(q_{X}, q_{Y}) \in \mathbb{R}_{+} \backslash \mathbb{R}_{++}$.

\subsubsection{Representative Inelastic Liquidity Trader (IT)}
The representative inelastic liquidity trader represents the liquidity traders who do not trade more when trading costs are lower. Liquidity traders may not trade more when trading costs decrease because they are otherwise constrained in trading, for example, because of a budget constraint, because of regulatory constraints, or because their demand for trading is satisfied regardless of trading costs. Therefore, we model the inelastic liquidity demand as a fixed constant, referred to as base demand, $B_{D}$. 

\subsection{Reverse Trade Arbitrage Opportunity}
We adopt the reverse trade arbitrage opportunity from \cite{capponi2021adoption}. After the representative elastic and inelastic liquidity traders have traded, the price implied by the AMM may differ from the fundamental exchange rate of the risky asset in terms of the \textit{numéraire}. This creates an arbitrage opportunity. Like the aforementioned arbitrageurs, we assume arbitrageurs have a perfect price signal and do not pay fees.

\section{Main Results}

Our solution concept is a pure-strategy subgame-perfect Nash equilibrium, which we find via backward induction. We start by setting a fixed amount of deposits from liquidity providers and find the best-response from the traders. Then we solve the subgame of competitive passive liquidity providers simultaneously providing optimal liquidity to maximize returns in anticipation of traders in future periods.

\begin{lemma} \label{demand_function_lemma}
    The expected aggregate demand of liquidity traders over $T$ periods in terms of the numéraire is
    $$
    D(d) = \frac{T}{\lambda} \Big( B_{D} + \frac{1}{2}(p_{0}q_{X}^{*} + q_{Y}^{*}) \Big),
    $$
    where
    $$
    (q_{X}^{*}, q_{Y}^{*}) = \begin{cases}
        (0, \enskip pd(\sqrt{\frac{\gamma_{U}}{1-f}} - \frac{1}{1-f}) & \text{if} \enskip p_{U} = \gamma_{U}p \\
        (d(\sqrt{\frac{\gamma_{U}}{1-f}} - \frac{1}{1-f}), \enskip 0) & \text{if} \enskip p_{U} = \gamma_{U}^{-1}p \\
    \end{cases}.
    $$
\end{lemma}

Lemma \ref{demand_function_lemma} shows that the expected aggregate demand of liquidity traders is increasing in the deposits of liquidity providers. As argued in \cite{capponi2024jit}, the $\gamma_{U}$ parameter shows how price elastic liquidity traders are.\\
The utility function of the $i$th liquidity provider is given as follows:

\begin{align} \label{LP_utility}
\Pi_{i} = \frac{d_{i}}{d} \Big( fD(d) - (A + r) \cdot V(p) \Big),
\end{align}
where $A$ is the arbitrage intensity per unit of liquidity expressed in terms of the numéraire, or the LVR \cite{milionis2024lvr}. $r$ is the opportunity cost of capital per unit of liquidity expressed in terms of the numéraire. $V(p) = 2pd$ is the pool value as a function of the external market price expressed in terms of the numéraire \cite{milionis2024lvr}.\\

If the arbitrage intensity and opportunity costs per unit of liquidity exceed the expected revenue from trading fees, liquidity providers will not deposit. If the expected marginal revenue from trading fees for one extra unit of liquidity always exceeds the arbitrage intensity and opportunity costs per unit of liquidity, then the liquidity providers will deposit their entire endowment as liquidity. In the remainder of this paper, we focus on the case in which the expected marginal revenue from trading fees for one extra unit of liquidity does not always exceed the arbitrage intensity and opportunity costs per unit of liquidity. This is the most relevant case as in the former case, there is a liquidity freeze \cite{capponi2024jit}, meaning that no assets are traded in the AMM. In the latter case, infinite amount of trading demand can be induced by supplying more liquidity. Since both of these situations lead to a trivial amount of deposits, we focus on the case in which the expected marginal revenue from trading fees for one extra unit of liquidity does not always exceed the arbitrage intensity and opportunity costs per unit of liquidity. Intuitively this is the case when the expected trading fees from the representative elastic liquidity demand are lower than the adverse selection and opportunity costs, but the combination with the representative inelastic liquidity demand is higher than the costs.

\begin{lemma} \label{total_deposits_lemma}
The total deposits of risky assets in the pool is given as follows:
$$
d^{*} = \frac{N-1}{N} \cdot \frac{f\frac{T}{\lambda}B_{D}}{2p_{0}(A+r) - f\frac{T}{\lambda} \frac{1}{2d^{*}}(p_{0}q_{X}^{*} + q_{Y}^{*})}.
$$
\end{lemma}

Lemma \ref{total_deposits_lemma} shows two important effects: the amount of deposits is increasing in the number of liquidity providers and is linearly increasing in the base demand. Furthermore, the total deposits converge to a constant as the number of liquidity providers goes to infinity. Now consider a monopolistic liquidity provider with the endowment of $N$ competitive liquidity providers. Since we focus on the case where the marginal revenue from an extra unit of liquidity from the representative elastic liquidity traders is smaller than the marginal costs for all amounts of liquidity, but the marginal revenue from an extra unit of liquidity from both the representative elastic and inelastic liquidity traders is larger than the marginal costs, the monopolistic liquidity provider aims to provide as little liquidity as possible to capture the base demand. We denote the smallest possible level of liquidity with $\epsilon$.

\begin{theorem} \label{poa_lps}
    The Price of Anarchy in the subgame of liquidity provision is linear in the number of liquidity providers.
    $$
    PoA_{LP} = \frac{\Pi_{mon}^{*}}{\Pi_{competitive}^{*}} = O(N).
    $$
    The expected aggregate profits of $N$ competitive liquidity providers, $\Pi_{competitive}^{*}$, and the expected profit of a monopolistic liquidity provider $\Pi_{mon}^{*}$ are given as follows:
    \begin{eqnarray}
        \Pi_{competitive}^{*} &=& \frac{1}{N} f \frac{T}{\lambda} B_{D}, \\
        \Pi_{mon}^{*} &=& f \frac{T}{\lambda} B_{D} + f \frac{T}{\lambda} \cdot 2p_{0}\epsilon \cdot \left(\sqrt{\frac{\gamma_{U}}{1-f}} - \frac{1}{1-f} - A - r\right).
    \end{eqnarray}
\end{theorem}

Theorem \ref{poa_lps} states that the price of anarchy of liquidity provider profits is linear in the number of liquidity providers. The expected aggregate profits of $N$ competitive liquidity providers depends on the fees gained from the representative inelastic liquidity traders and the number of liquidity providers competing for these fees. The monopolist LP, however, can extract all fee revenue from the inelastic liquidity demand. This means that the $N$ competitive liquidity providers could be better off by collaborating and solely capturing the base demand instead of engaging in a costly competition of providing more liquidity to extract fee revenue from the base demand even though the additional liquidity does not induce more inelastic liquidity demand. Theorem \ref{poa_lps} shows a disadvantage of the pro-rata allocation rule. The pro-rata rule forces competitive liquidity providers to compete for fee revenues by putting more capital at risk of adverse selection and incurring more opportunity costs. Although the amount of liquidity in the pool is increasing in the number of liquidity providers, the utility of liquidity providers is decreasing in the number of liquidity providers and therefore also in the amount of liquidity in the pool.

The utility of the representative elastic liquidity trader is given as follows:
\begin{equation}
\Pi_{ET} = \frac{1}{2}pd \cdot \Gamma(\gamma_{U}, f),
\end{equation}
where $\Gamma(\gamma_{U},f) = 1 + \gamma_{U} - \frac{4}{\sqrt{\gamma_{U}(1-f)}} + \frac{1}{1-f}(\frac{1}{\gamma_{U}} + 1).$
The utility of the representative elastic liquidity trader is linearly increasing in the deposits in the pool and depends on the private valuation and trading fees. $\Gamma(\gamma_{U},f)$ can be interpreted as a measure of elasticity of demand.

\begin{theorem} \label{poa_full_game}
    The Price of Anarchy of the representative elastic liquidity traders and the liquidity providers is given as follows
    $$
    PoA = O(1).
    $$
\end{theorem}

Theorem \ref{poa_full_game} shows that the welfare loss that Theorem \ref{poa_lps} showed scales linearly in the number of liquidity providers in the subgame of liquidity provision disappears when considering the utility of the representative elastic liquidity trader. This means that the representative inelastic liquidity trader and the elastic liquidity trader complement each other. Since the competitive liquidity providers want to capture the fee revenue of the inelastic liquidity trader, as shown in Theorem \ref{poa_lps}, the LPs provide a lot of capital in order to increase their individual portion of fee revenue. This increase in liquidity leads to a higher welfare for elastic liquidity traders. Potentially, another fee allocation rule would not exploit this complementarity. It is not clear, however, if this complementarity could be considered fair or efficient. This would warrant further research into the utility of the inelastic base demand. This paper shows that understanding the different utility functions of different types of liquidity traders is essential to understanding whether a certain allocation rule is desirable.

\section{Loss-Versus-Rebalancing Rebating}

We now consider an AMM that rebates a fraction $k \in [0,1]$ of the adverse selection costs that liquidity providers incur. This is a form of rebating LVR \cite{milionis2024fees}, which is a big line of focus for new AMM designs. In this section, we research what the effect is of a fractional rebate of LVR on the utilities of liquidity providers and the representative elastic liquidity trader.

We modify Equation \ref{LP_utility} to include a rebate parameter.
\begin{equation}
    \Pi_{i} = \frac{d_{i}}{d}\Big(f D(d) - ((1-k)A + r) \cdot V(p)\Big),
\end{equation}
First, we find how the equilibrium deposits are affected by the rebate parameter. Notice that the condition for non-trivial liquidity provision is now given as $f\frac{dD(d)}{dd} < 2p_{0}((1-k)A + r) < f\Big( \frac{dD(d)}{dd} + \frac{T}{\lambda} B_{D} \Big)$.

\begin{corollary} \label{rebate_parameter_lemma}
The partial derivative of the equilibrium deposits in the presence of LVR rebate with respect to the rebate parameter is given as follows.
    \begin{align*}
    \frac{\partial d^{*}}{\partial k} = \frac{\frac{N-1}{N} \cdot f \cdot \frac{T}{\lambda} B_{D} \cdot 2p_{0}A}{(2p_{0}((1-k)A + r) - f \frac{T}{\lambda} \frac{1}{2d^{*}}(p_{0}q^{*}_{X} + q^{*}_{Y}))^{2}} > 0.
    \end{align*}
    The utility of the representative elastic liquidity trader therefore has the following relationship with the rebate parameter
    \begin{align*}
    \frac{\partial \Pi_{ET}}{\partial k} = \frac{1}{2}p \frac{\partial d^{*}}{\partial k} \Gamma(\gamma_{U}, f) > 0.
    \end{align*}
\end{corollary}

From Lemma \ref{rebate_parameter_lemma}, it follows that the equilibrium deposits are increasing in the rebate parameter and therefore the utility of the representative elastic liquidity trader is increasing in the rebate parameter.

\begin{corollary} \label{no_rebate_LPs}
    The profits of competitive liquidity providers are not affected by rebates of adverse selection costs if 
    \begin{align*}
        f\frac{dD(d)}{dd} < 2p_{0}((1-k)A + r) < f\Big( \frac{dD(d)}{dd} + \frac{T}{\lambda} B_{D} \Big) \enskip \text{for every} \enskip k \in [0,1].
    \end{align*}
\end{corollary}

Corollary \ref{no_rebate_LPs} states that liquidity providers do not gain in utility when a rebate parameter is implemented. This is counterintuitive, as much of recent AMM design has focused on rebating LVR because LPs would suffer too much adverse selection costs. Corollary \ref{no_rebate_LPs} follows immediately from adapting Equation \ref{LP_utility} and Theorem \ref{poa_lps}. 

\begin{lemma} \label{rebate_lemma}
    A decrease of the arbitrage intensity per unit of liquidity of $k$ leads to a less than proportional decrease of the total arbitrage intensity
    \begin{align*}
    \frac{(1-k)AV^{rebate}}{AV^{no-rebate}} \geq 1-k,
    \end{align*}
    where $V^{rebate}$ is the equilibrium pool value with a rebate of $k$ and $V^{no-rebate}$ is the equilibrium pool value without a rebate.
\end{lemma}

Lemma \ref{rebate_lemma} shows that the total arbitrage intensity decreases less than proportionally when the arbitrage intensity per unit of liquidity decreases by a fraction of $k$. This implies that AMM designs aimed at reducing the arbitrage intensity per unit of liquidity do not lead to a proportional reduction of adverse selection costs. In fact, the total arbitrage amount could even increase if the equilibrium amount of deposits increases sufficiently in the case of a rebate compared to no rebate. Potentially MEV mitigations at the application layer do not prevent the centralization concerns that arise at the protocol layer. This may warrant more research into whether applayer MEV minimizations can prevent MEV centralization concerns at the protocol layer.
\bibliographystyle{splncs04}
\bibliography{references}
\section{Appendix}

\subsection{Proof of Lemma \ref{demand_function_lemma}}
Blocks are generated by a Poisson process with rate $\lambda$ during an interval $[0,T]$. In each block, the representative inelastic liquidity trader trades $B_{D}$ units of assets in terms of the numéraire. In each block the representative elastic liquidity trader receives a private valuation $p_{U}$ and trades to maximize its utility.

Following \cite{capponi2024jit}, we define $\delta_{Y}(x, d, f)$ to be the number of numéraire that $x$ risky assets can be swapped for at pool deposits of $d$ and a trading fee of $f$. Similarly, let $\delta_{X}(y, d, f)$ be the number of risky assets that $y$ numéraire can be swapped for at pool deposits of $d$ and a trading fee of $f$.

\begin{align*}
    \delta_{Y}(x,d,f) = pd - \frac{pd^{2}}{d + (1-f)x} = \frac{pd(1-f)x}{d + (1-f)x} \\
    \delta_{X}(y,d,f) = d - \frac{pd^{2}}{pd + (1-f)y} = \frac{d(1-f)y}{pd + (1-f)y}
\end{align*}

The utility of the ET is given
$$
u_{ET} = \begin{cases}
    p_{U} \cdot \delta_{X}(y, d, f) - y & \text{if} \enskip p_{U} = \gamma_{U}p \\
    \delta_{Y}(x, d, f) - p_{U} \cdot x & \text{if} \enskip p_{U} = \gamma_{U}^{-1}p \\
\end{cases}
$$

It then follows that the optimal trade for a ET is given as follows:
$$
(q_{X}^{*}, q_{Y}^{*}) = \begin{cases}
    (0, pd(\sqrt{\frac{\gamma_{U}}{1-f}} - \frac{1}{1-f})) & \text{if} \enskip p_{U} = \gamma_{U}p \\
    (d(\sqrt{\frac{\gamma_{U}}{(1-f)}} - \frac{1}{1-f}), 0) & \text{if} \enskip p_{U} = \gamma_{U}^{-1}p \\
\end{cases}
$$
Therefore, the expected aggregate trading demand in terms of the numéraire over the interval $[0,T]$ is given as follows:
$$
D(\gamma_{H}, d) = \frac{T}{\lambda} \Big(B_{D} + \frac{1}{2}(p_{0}q_{X}^{*} + q_{Y}^{*}) \Big)
$$
Arbitrageurs are excluded from the expected aggregate trading demand as they do not pay fees.

\subsection{Proof of Lemma \ref{total_deposits_lemma}}

The utility of liquidity providers is given in Equation \ref{LP_utility}. Maximizing the utility of the $i$th LP with respect to its deposits given the deposits of the other LPs yields the following best-response function.
$$
d_{i} = \sqrt{\frac{\sum_{j \in [N], j \neq i} d_{j} f\frac{T}{\lambda} B_{D}}{2p(A+r) - f\frac{T}{\lambda}\frac{1}{2d}(q_{X}^{*} + q_{Y}^{*})}} - \sum d_{j}
$$
We then find the dominant strategy for all LPs, $d_{i}^{*}$, by symmetry:
$$
d_{i}^{*} = \frac{N-1}{N^{2}} \cdot \frac{\frac{T}{\lambda}B_{D}}{2p(A + r) - \frac{fT}{\lambda}\frac{1}{2d}(q_{X}^{*} + q_{Y}^{*})}
$$
The total deposits and pool value \cite{milionis2024fees} in equilibrium are given as follows.
$$
d^{*} = Nd_{i}^{*} \qquad V^{*}(p) = 2pd^{*}
$$

\subsection{Proof of Theorem \ref{poa_lps}}
We first find the profits of $N$ liquidity providers when there is competition and when there is a monopolist LP with the endowment of $N$ LPs separately.

By substituting the equilibrium deposits from Lemma \ref{total_deposits_lemma}, we find the profits for an individual LP in competition.
$$
\Pi_{i} = \frac{1}{N^{2}} f\frac{T}{\lambda} B_{D}
$$
The total LP profits in competition are then given as follows
$$
\Pi^{comp} = N \cdot \Pi_{i} = \frac{1}{N} f\frac{T}{\lambda} B_{D}
$$
Consider a monopolistic liquidity provider who receives an endowment of $(Ne_{X}, Np_{0}e_{Y})$, hence the equivalent of $N$ competitive LPs. The deposits of the monopolist LP are then given by
$$
d^{mon} = \begin{cases}
    Ne_{X} & \text{if} \enskip 2p(A+r) \leq \frac{dD(d^{mon})}{dd^{mon}} \\
    \epsilon & \text{if} \enskip \frac{dD(d^{mon})}{dd^{mon}} < 2p(A + r) \leq fD(\epsilon) \\
    0 & \text{if} fD(\epsilon) < 2p(A + r)
\end{cases}
$$
Our analysis focuses on the second case. This case corresponds to a situation in which the marginal induced demand of the representative elastic and inelastic liquidity trader of one unit additional of liquidity is larger than the marginal adverse selection and opportunity costs, but the marginal induced demand of the representative elastic liquidity trader is smaller than the marginal adverse selection and opportunity costs. Therefore, we find the following expression for the expected profit of the monopolist LP.
$$
\Pi^{mon} = f \frac{T}{\lambda}B_{D} + f\frac{T}{\lambda}p_{0}\epsilon(\sqrt{\frac{\gamma_{U}}{1-f}} - \frac{1}{1-f}) - 2p_{0}\epsilon(A + r)
$$

Consider $c = \frac{\Pi^{mon}}{f \frac{T}{\lambda}B_{D}}$ and $k = 1$, we then find the following
$$
0 \leq PoA = \frac{\Pi^{mon}}{\Pi^{comp}} \leq c \cdot N
$$
Hence we have the final result
$$
PoA = O(N)
$$

\subsection{Proof of Theorem \ref{poa_full_game}}
Consider the price of anarchy of the representative elastic liquidity trader and the liquidity providers in the full-game.
$$
PoA = \frac{\Pi^{mon} + \frac{1}{2}p\epsilon \cdot \Gamma(\gamma_{U}, f)}{\Pi^{comp} + \frac{1}{2}pd^{*} \cdot \Gamma(\gamma_{U},f)}
$$
We know $f\frac{T}{\lambda}B_{D} > p_{0}(2(A+r) - f \frac{T}{\lambda}(\sqrt{\frac{\gamma_{U}}{1-f}} - \frac{1}{1-f})) $. Let $z$ denote the variable such that $f\frac{T}{\lambda}B_{D} = z \cdot p_{0}(2(A+r) - f \frac{T}{\lambda}(\sqrt{\frac{\gamma_{U}}{1-f}} - \frac{1}{1-f})) $. We then denote the price of anarchy as
$$
PoA = \frac{\Pi^{mon} + \frac{1}{2}p\epsilon \Gamma(\gamma_{U}, f)}{\frac{1}{2}pz\Gamma(\gamma_{U}, f) + \frac{1}{N}(f\frac{T}{\lambda}B_{D} - \frac{1}{2}pz \Gamma(\gamma_{U},f))}
$$
If $f\frac{T}{\lambda}B_{D} \geq \frac{1}{2}pz \Gamma(\gamma_{U},f)$, then $\frac{1}{2}pz\Gamma(\gamma_{U}, f) + \frac{1}{N}(f\frac{T}{\lambda}B_{D} - \frac{1}{2}pz\Gamma(\gamma_{U}, f)) \geq \frac{1}{2}pz\Gamma(\gamma_{U}, f)$, hence we find the price of anarchy as follows
$$
PoA \leq \frac{\Pi^{mon} + \frac{1}{2}p\epsilon \Gamma(\gamma_{U},f)}{\frac{1}{2}pz\Gamma(\gamma_{U},f)} = O(1)
$$
If $f \frac{T}{\lambda} B_{D} < \frac{1}{2}pz\Gamma(\gamma_{U},f)$, then $\frac{1}{2}pz\Gamma(\gamma_{U}, f) + \frac{1}{N}(f\frac{T}{\lambda}B_{D} - \frac{1}{2}pz\Gamma(\gamma_{U}, f)) \geq f \frac{T}{\lambda} B_{D}$, then it follows:
$$
PoA \leq \frac{\Pi^{mon} + \frac{1}{2}p\epsilon \Gamma(\gamma_{U}, f)}{\frac{1}{2}pz\Gamma(\gamma_{U}, f)} = O(1)
$$
Hence, the price of anarchy of the representative elastic liquidity trader and the LPs is $PoA = O(1)$.

\subsection{Proof of Lemma \ref{rebate_lemma}}

The proof immediately follows given that Corollary \ref{rebate_parameter_lemma} shows that the total deposits are increasing in the rebate parameter, hence $V^{rebate} > V^{no-rebate}$, hence $\frac{(1-k)AV^{rebate}}{AV^{no-rebate}} \geq 1-k$.

Notice that there is an increase in total arbitrage if $\frac{V^{rebate}}{V^{no-rebate}} \geq \frac{1}{1-k}$ or when $V^{rebate}(1-k) \geq V^{no-rebate}$.
\end{document}